\documentclass[fleqn,usenatbib]{mnras}

\DeclareRobustCommand{\VAN}[3]{#2}
\let\VANthebibliography\thebibliography
\def\thebibliography{\DeclareRobustCommand{\VAN}[3]{##3}\VANthebibliography}

\defcitealias{Noon:2023aa}{N23}
\defcitealias{Gerrard:2023aa}{G23}
\usepackage[T1]{fontenc}
\usepackage{graphicx}	% Including figure files
\usepackage{amsmath}	% Advanced maths commands
\usepackage{amssymb}	% Extra maths symbols
\usepackage{amsfonts}
\usepackage{caption}
\usepackage{subcaption}
\usepackage{xcolor}

\usepackage{scalerel,tikz}
\usetikzlibrary{svg.path}
\definecolor{orcidlogocol}{HTML}{A6CE39}
\tikzset{orcidlogo/.pic={\fill[orcidlogocol] svg{M256,128c0,70.7-57.3,128-128,128C57.3,256,0,198.7,0,128C0,57.3,57.3,0,128,0C198.7,0,256,57.3,256,128z}; \fill[white] svg{M86.3,186.2H70.9V79.1h15.4v48.4V186.2z} svg{M108.9,79.1h41.6c39.6,0,57,28.3,57,53.6c0,27.5-21.5,53.6-56.8,53.6h-41.8V79.1z M124.3,172.4h24.5c34.9,0,42.9-26.5,42.9-39.7c0-21.5-13.7-39.7-43.7-39.7h-23.7V172.4z} svg{M88.7,56.8c0,5.5-4.5,10.1-10.1,10.1c-5.6,0-10.1-4.6-10.1-10.1c0-5.6,4.5-10.1,10.1-10.1C84.2,46.7,88.7,51.3,88.7,56.8z};}}
\newcommand\orcidicon[1]{\href{https://orcid.org/#1}{\mbox{\scalerel*{
\begin{tikzpicture}[yscale=-1,transform shape]\pic{orcidlogo};
\end{tikzpicture}}{|}}}}

\newcommand{\msun}{\mathrm{M}_{\odot}}

\newcommand{\mach}{\mathcal{M}}
\newcommand{\sigr}{\sigma_{\rho/\rho_0}}
\newcommand{\sigv}{\sigma_{v,\mathrm{3D}}}

\newcommand{\pthreed}{P_\mathrm{3D}}
\newcommand{\ptwod}{P_\mathrm{2D}}
\newcommand{\cs}{c_\mathrm{s}}
\newcommand\hi{\mbox{\sc Hi}}

\newcommand{\brunt}{\mathcal{R}^{1/2}}
\newcommand{\hm}{\mathrm{H}_{2}}
\newcommand{\minerr}{16^{\mathrm{th}}}
\newcommand{\maxerr}{84^{\mathrm{th}}}
\title[]{Turbulence statistics of $\hi$ clouds entrained in the Milky Way's nuclear wind}

\author[Gerrard et al.]{
Isabella~A.~Gerrard$^{\orcidicon{0000-0002-1995-6198}\,1}$\thanks{E-mail: isabella.gerrard@anu.edu.au}, Karlie~ A.~Noon$^{\orcidicon{0000-0002-9699-6863}\,1}$, Christoph~Federrath$^{\orcidicon{0000-0002-0706-2306}\,1,2}$, Enrico~M.~Di~Teodoro$^{\orcidicon{0000-0003-4019-0673}\,3}$, 
\newauthor\hspace{0.02cm}
Antoine~Marchal${\orcidicon{0000-0002-5501-232X}\,^1}$, N.~M.~McClure-Griffiths${\orcidicon{0000-0003-2730-957X}\,^1}$, 
\\
$^{1}$Research School of Astronomy and Astrophysics, Australian National University, ACT 2611, Australia\\
$^{2}$Australian Research Council Centre of Excellence in All Sky Astrophysics (ASTRO3D), Canberra, ACT 2611, Australia\\
$^{3}$Dipartimento di Fisica e Astronomia, Universit\`a degli Studi di Firenze, via G. Sansone 1, 50019 Sesto Fiorentino, Firenze, Italy\\
}
\date{Accepted XXX. Received YYY; in original form ZZZ}

\pubyear{2024}

\begin{document}
\label{firstpage}
\pagerange{\pageref{firstpage}--\pageref{lastpage}}
\maketitle

\begin{abstract}
The interstellar medium (ISM) is ubiquitously turbulent across many physically distinct environments within the Galaxy. Turbulence is key in controlling the structure and dynamics of the ISM, regulating star formation, and transporting metals within the Galaxy. We present the first observational measurements of turbulence in neutral hydrogen entrained in the hot nuclear wind of the Milky Way. Using recent MeerKAT observations of two extra-planar \hi{} clouds above (gal.\,lat.$\,\sim7.0^{\circ}$) and below (gal.\,lat.$\,\sim-3.9^{\circ}$) the Galactic disc, we analyse centroid velocity and column density maps to estimate the velocity dispersion ($\sigv$), the turbulent sonic Mach number ($\mach$), the volume density dispersion ($\sigr$), and the turbulence driving parameter ($b$). We also present a new prescription for estimating the spatial temperature variations of $\hi$ in the presence of related molecular gas. We measure these turbulence quantities on the global scale of each cloud, but also spatially map their variation across the plane-of-sky extent of each cloud by using a roving kernel method. We find that the two clouds share very similar characteristics of their internal turbulence, despite their varying latitudes. Both clouds are in the sub-to-trans-sonic Mach regime, and have primarily compressively-driven ($b\sim1$) turbulence. Given that there is no known active star-formation present in these clouds, this may be indicative of the way the cloud-wind interaction injects energy into the entrained atomic material on parsec scales.                                            
\end{abstract}


%%%%%%%%%%%%%%%%%%%%%%%%%%%%%%%%%%%%%%%%%%%%%%%%%%
\section{Introduction}
%%%%%%%%%%%%%%%% turbulence/ISM:
Turbulence is one of the main driving forces that shapes the evolution of the interstellar medium (ISM) in galaxies. To understand the mechanisms governing the life cycle of the ISM, we need to characterise the turbulence we observe in a diverse range of galactic environments through comparison with theoretical models and simulations \citep[see review by][]{Burkhart2021}. Statistical methods such as the Spatial Power Spectrum (SPS) \citep[e.g.][]{Stanimirovic:1999aa, Stanimirovic:2001aa, KowalLazarian2007, HeyerEtAl2009, Chepurnov:2015aa, Nestingen-Palm:2017aa, Pingel:2018aa, Szotkowski:2019wa} and the Probability Distribution Function (PDF) \citep[e.g.,][]{Burkhart:2010aa, Patra:2013aa, Bertram:2015aa, Nestingen-Palm:2017aa, Maier:2017aa} are important because they allow us to compare the distinctly two-dimensional (2D) information we can measure on the plane of the sky with the three-dimensional (3D) information we can access in increasingly sophisticated (magneto)hydrodynamic (MHD) simulations of the ISM \citep[e.g.,][]{Kim:2017aa,Kim:2018ab,Kim:2023aa} 

The density PDF of simulated isothermal, supersonic gas can be described by a log-normal function, meaning that the logarithm of the density follows a normal Gaussian distribution \citep{Vazquez-Semadeni:1994vq, PadoanJonesNordlund1997, Passot:1998wz, FederrathKlessenSchmidt2008,Hopkins2013PDF,SquireHopkins2017,BeattieEtAl2021}. The fluctuations of the density field are captured in the width of this PDF, a key parameter in star-formation theories \citep{KrumholzMcKee2005, PadoanNordlund2011, HennebelleChabrier2011, FederrathKlessen2012,BurkhartMocz2019}. Furthermore, studies by \citet{PriceFederrathBrunt2011,KonstandinEtAl2012ApJ,MolinaEtAl2012,NolanFederrathSutherland2015,FederrathBanerjee2015,KainulainenFederrath2017} have shown that the width of the density PDF is proportional to the turbulent (sonic) Mach number. Formally, this is described by
\begin{equation} \label{eq:b}
    \sigr = b\mach,
\end{equation}
where $\sigr$ is the standard deviation of the PDF of the 3D density field ($\rho$), scaled by the mean density ($\rho_0$), $\mach$ is the turbulent sonic Mach number, and $b$ is a constant of proportionality known as the \emph{turbulence driving parameter} \citep{FederrathKlessenSchmidt2008,Federrath:2010wz}. This constant of proportionality is another key ingredient in star-formation theories, and describes the ratio of compressive and solenoidal modes in the \textit{acceleration field} of the gas, which is what drives the turbulence. We refer the reader to the large body of work testing the theoretically-derived $b$ parameter in numerical studies of driven turbulence \citep[e.g.][]{FederrathKlessenSchmidt2008,Federrath:2010wz,PriceFederrathBrunt2011,MolinaEtAl2012,NolanFederrathSutherland2015,FederrathBanerjee2015,BeattieEtAl2021}, as well as to a more in-depth discussion of the functionality of $b$ in an observational context in \citet{Gerrard:2023aa} (hereafter: \citetalias{Gerrard:2023aa}). \citetalias{Gerrard:2023aa} described a new method for extracting $\sigr$, $\mach$, $b$ (which we refer to as the main analysis quantities within this text) from observations, specifically from $\hi$ observations of the SMC. 

Recent high-resolution observations of the large-scale wind activity in the centre of the Milky Way has revealed significant amounts of molecular and atomic gas entrained in the wind \citep[][hereafter: \citetalias{Noon:2023aa}]{Di-Teodoro:2018aa,Di-Teodoro:2020aa, Noon:2023aa}. In this study, we will use the \citetalias{Gerrard:2023aa} techniques to probe the internal turbulence of two such extra-planar $\hi$ clouds. Extending our analysis to extra-planar clouds, positioned at significant angular distances from the Galactic plane, introduces an additional layer of complexity to our understanding of interstellar dynamics. These clouds present unique laboratories for studying turbulence in the ISM without the influence of ongoing internal star formation activity as is present in most cold clouds in the Galactic disc.

Theoretical studies encounter difficulties when attempting to simulate a fully entrained cold gas cloud in a hot wind before instabilities destroy the cloud \citep{Zhang:2017aa, Schneider:2017aa, Girichidis:2021aa}, whilst other simulations find that it might be possible given specific criteria such as a large initial cloud mass or molecular gas forming via self-shielding in an entrained atomic cloud \citep{Armillotta:2016aa, Armillotta:2017aa, Gronke:2018aa, Farber:2022aa, Chen:2023aa}. Observational analysis of three such clouds has revealed that they were possibly molecular-dominated in the disc, but have since been accelerated in the nuclear wind, triggering the photo-dissociation of the molecular gas \citepalias{Noon:2023aa}. \citet{Di-Teodoro:2018aa} and \citet{Lockman:2020aa} use a kinematic, bi-conical wind model to determine the distances to each cloud from the Galactic centre (GC), revealing that the intermediate-latitude cloud also possesses broader CO line profiles, a disordered velocity field and a wider FWHM range than the closest cloud. Further, the most distant cloud possesses no detectable molecular gas. The differences in cloud properties as a function of distance could indicate a cloud's evolution in the Galactic wind, suggesting they slowly lose molecular content as they transit the wind. Alternatively, it could indicate varying localised conditions in the wind or stem from the relationship between cloud properties and survival in a wind. Probing the internal turbulence of these clouds could further reveal their history. Our aim is to see whether the cloud's turbulence characteristics vary between the clouds or with distance from the GC.

The rest of this work is structured as follows: in Sec.~\ref{sec:methods} we outline our data preparation and the modifications we have made to the \citetalias{Gerrard:2023aa} methods, in Sec.~\ref{sec:results} we discuss the results of our analysis and present some potential interpretations. We summarise our results in Sec.~\ref{sec:summary}.

%%%%%%%%%%%%%%%%%%%%%%%%%%%%%%%%%%%%%%%%%%%%%%%%%%
\section{Methods}\label{sec:methods}
In this section we provide an overview of the methods used to extract and map the turbulent properties of the $\hi$ gas in two extra-planar clouds. Specifically, we recover the main analysis quantities $\sigr$, $\mach$, $b$, as well as the intermediate analysis quantities $\sigv$, $\brunt$ and $\cs$ (described in full in the following sections). Details of the methodology are outlined in \citetalias{Gerrard:2023aa}. For the present work, we have made some modifications to the method so that the pipeline can analyse data cubes that have a significant amount of empty pixels (see Sec.~\ref{sssec:kernel}), as well as adding a low-pass filtering method (further discussed in Sec.~\ref{sssec:LPF}).

\subsection{Data}\label{ssec:obs}
Here we will outline the characteristics of the data relevant to the analysis presented in this paper. For a more detailed summary on the observation set up and data reduction, see \citet{Di-Teodoro:2020aa} and \citetalias{Noon:2023aa}. We analyse two of the three extra-planar clouds presented in \citetalias{Noon:2023aa}, C1 (closest to the GC) and C3 (furthest from the GC). The intermediate-latitude cloud, C2, has a particularly complex velocity structure, as is likely a conglomerate of several distinct structures coincident in position-position-velocity (PPV) space. Our methods are not applicable to this kind of object, and as such, we have excluded it from the analysis.

We observed \hi{} 1.42\,GHz and $^{12}$CO(2$\rightarrow$1) 230.538\,GHz emission lines using the MeerKAT radio interferometer in late 2020 (SARAO project DDT-20201006-NM-01; principal investigator N.M.G) and the Atacama Pathfinder EXperiment (APEX) telescope (ESO project 0104.B-0106A; principal investigator E.M.D.T.), respectively. The \hi{} observations have a spectral resolution of 5.5\,km\,s$^{-1}$ and a spatial resolution of 24.9$''$ $\times$ 21.7$''$ with position angle $-2.8^{\circ}$. The CO observations have a spectral resolution of 0.2\,km\,s$^{-1}$ and a spatial resolution of 28$''$. The corresponding root-mean-square (rms) noise in the \hi{} data cubes ($\sigma_{\mathrm{chn}}$) are 304\,mK and 365\,mK for C1 and C3, respectively, per 5.5\,km\,s$^{-1}$ channel. The rms noise of the CO data cubes is 65\,mK per 0.25\,km\,s$^{-1}$ channel for both clouds.

\subsection{Data preparation}\label{ssec:data_prep}
\begin{figure*}
     \centering
     \includegraphics[width=\linewidth]{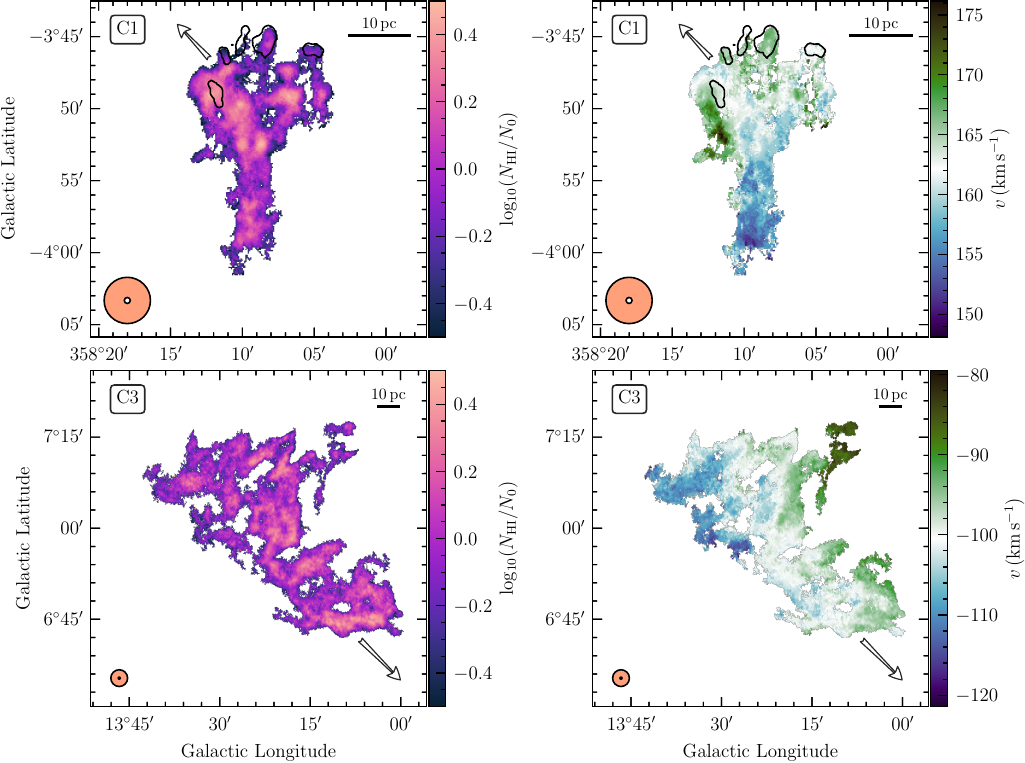}
     \caption{Moment maps for each of the clouds analysed in this work. The left-hand panels show the logarithmic normalised column density, while the right-hand panels show the centroid velocity, with C1 on the top and C3 on the bottom. The orange circle shows the FWHM of the kernel, while the small white circle represents the beam of the \hi{} observations. The black contours in the top panels show $\hm$ column density at $5.0\times10^{21}$\,cm$^{-2}$. The scale bar shows 10\,pc, assuming the physical distance to the clouds is 8.3\,kpc. In each panel, the arrow points toward the Galactic Center (GC), indicating the opposite direction of the wind's flow.}
     \label{fig:moments_maps}
\end{figure*}
Before inputting the column density and centroid velocity maps of each cloud to the \citetalias{Gerrard:2023aa} pipeline, we first perform a few data-cleaning processes. The first is to make a per-channel signal-to-noise (SNR) cut at 3$\sigma_{\mathrm{chn}}$. This removes much of the low-SNR emission from around the central bulk of the cloud, as well as spurious signals. We then create the moment-0 map ($\mathrm{M}_0$), given by
\begin{equation} \label{eq:M0}
\mathrm{M_0}(x,y) = \int T_b(x,y,v)\,dv, 
\end{equation}
where $T_b(x,y,v)$ is the brightness temperature (K) in the velocity channel $v$ and in the pixel $(x,y)$, and $dv$ is the channel spacing (km\,s$^{-1}$). We calculate the column density in cm$^{-2}$ of each cloud by multiplying the $\mathrm{M}_0$ by a factor of $1.823 \times 10^{18}$ cm$^{-2}$~(K\,km\,s$^{-1}$)$^{-1}$, assuming the gas is optically thin \citep{McClure-Griffiths:2023aa}.

The moment-1 ($\mathrm{M}_1$) is the intensity-weighted centroid velocity, and is given by
\begin{equation} \label{eq:M1}
\mathrm{M_1}(x,y) = \frac{\int v\,T_b(x,y,v)\,dv}{\int T_b(x,y,v)\,dv} =\frac{\int v\,T_b(x,y,v)\,dv}{\mathrm{M_0}}, 
\end{equation}
where $v$ is the velocity of each channel. 

Post integration, we construct a map of the noise in the integrated intensity map to perform another SNR cut. We follow the method outlined in \citet{Lelli:2014aa} (Appendix~A1; Eq.~A5) for a uniform tapered cube that has been continuum-subtracted, defining
\begin{equation} \label{eq:rms_m0}
\sigma_{\mathrm{M0}} = \sqrt{1+0.25 N_{\mathrm{chn}} (1/N_1 + 1/N_2)}\times\sqrt{N_{\mathrm{chn}}}\times\sigma_{\mathrm{chn}}dv,
\end{equation}
where $\sigma_{\mathrm{M0}}$ is the noise in the integrated map, $\sigma_{\mathrm{chn}}$ is the intrinsic noise per channel, $N_{\mathrm{chn}}$ is the number of channels integrated to make the moment-0 map, and $N_1$ and $N_2$ are the number of channels used during continuum subtraction from each end of the spectral axis of the cube ($N_1 = 86, 133$ and $N_2 = 96, 49$ for C1 and C3 respectively). Once we have constructed a map of the noise, we create a SNR map, and use it to threshold the moment-0 map. We choose to include everything above 5$\sigma_{\mathrm{M0}}$ in the maps for the following analysis.

Lastly, we use a watershed algorithm \citep{fiorio:inria-00549539, Wu:2005aa} to find the largest region connected pixel-by-pixel in the moment-0 map and mask out any remaining pixels outside of that region, so as to exclude any small `islands' of emission. This mask is also applied to the moment-1 maps.

\subsubsection{Roving kernel}\label{sssec:kernel}
As in \citetalias{Gerrard:2023aa}, we use a roving kernel to map the turbulent quantities across the on-sky extent of the clouds. This is a Gaussian kernel, truncated at 3~sigma. The method in \citetalias{Gerrard:2023aa} only computes the analysis quantities in kernels which are entirely filled with valid data, and assigns NaN to any kernel which has empty pixels inside of it. In this new iteration of the method, kernel windows are permitted to contain empty pixels, but must have a minimum number of valid pixels contained within the kernel FWHM to be processed. In essence, this allows us to compute statistics in kernels that contain a few NaN pixels, such as at the edges of the clouds, while still maintaining confidence in those statistics. As described in \citet{Sharda:2018tk}, the minimum number of resolution elements required to recover a reasonably converged standard deviation is of order 20, so we set a minimum of 20~beams worth of viable pixels inside the kernel's FWHM for that kernel to be processed. We chose the actual size of the kernel such that the FWHM is the diameter of a circle that has an area 3 times larger than the minimum number of valid pixels. This is discussed further in Appendix~\ref{app:kernel_size}. This change to the \citetalias{Gerrard:2023aa} method allows us to analyse objects with a comparable number of independent resolution elements to the minimum size of the kernel FWHM, which was not necessary in the original work because the kernel was small compared to the entire SMC data cube and the number of beams contained within it.

\subsubsection{Low-pass filtering}\label{sssec:LPF}
In addition to the kernelled maps that allow us to spatially resolve turbulent properties of the clouds, we also compute global statistics for each cloud, meaning that we calculate the main and intermediate analysis quantities using the entire cloud to measure the variance in density and velocity. This allows us to make a comparison of the properties on two different spatial scales: the minimum size that the variance can be resolved in, and the largest scale, i.e. the entire cloud. In order to isolate the purely turbulent fluctuations in both density and velocity on the whole-cloud scale, we perform a Gaussian convolution of the column density and centroid velocity maps, to create a low-pass filter (LPF) map that we then subtract from the original maps. By using this convolution method to construct the large-scale variations across each cloud, we can account for their unique geometry and clumpy $\hi$ emission structure. The size of the convolution kernel is dictated by the size of the cloud: we take the total area of the cloud and use the radius of a circle with that same area as the FWHM of the Gaussian. We use these LPF-corrected maps to compute the global analysis quantities of each cloud, as well as the input to the roving kernel method, although for continuity of the method we still perform a linear gradient subtraction on the kernel scale. Whether or not we pass the LPF-corrected maps or the raw moment-0 and moment-1 maps does not effect the results of the kernelling method, however. In Appendix~\ref{app:LPF} the results of the LPF-method can be seen for both density and velocity, for each cloud.

\subsection{Turbulence analysis}\label{ssec:pipeline}
Our approach to extracting the density and velocity statistics for each cloud centres around isolating the turbulent fluctuations in the column density and centroid velocity maps, and using these plane-of-sky quantities to reconstruct the associated three-dimensional (3D) quantities required for estimating the volume density dispersion, the 3D velocity dispersion, the Mach number and the turbulence driving parameter for each cloud as per Eq.~\ref{eq:b}. We compute these quantities in a roving Gaussian kernel, where we subtract a kernel-scale gradient from the LPF-corrected moment-0 and moment-1 maps, so as to isolate the turbulent fluctuations before measuring the plane-of-sky dispersion. An in-depth description of the pipeline can be found in \citetalias{Gerrard:2023aa}, but the main points are summarised in the following and changes with respect to \citetalias{Gerrard:2023aa} highlighted where applicable.

\subsubsection{Density statistics}\label{sssec:dens_stats}
To convert the plane-of-sky column density dispersion $\sigma_{N_{\mathrm{HI}}/N_0}$ to the volume density dispersion, $\sigr$, we follow the method outlined in \cite{BruntFederrathPrice2010a}. The relation between the 2D density power spectrum $\ptwod(k)$ and the 3D density power spectrum $\pthreed(k)$ is given by
\begin{equation}
    \pthreed(k) = 2k \ptwod(k),
    \label{eqn:P3dP2d}
\end{equation}
where $k$ is the wave number \citep{Federrath:2013wt}. We exploit this relation to recover $\pthreed(k)$ by first computing $\ptwod(k)$ of the gradient-subtracted column density, which immediately gives us $\pthreed(k)$ of the quantity $\rho/\rho_0 - 1$ as per Eq.~(\ref{eqn:P3dP2d}). The ratio of the sums over $k$-space of these two quantities gives the density variance ratio \citep{BruntFederrathPrice2010a},
\begin{equation}
    \mathcal{R}^{1/2} = \frac{\sigma_{N_{\mathrm{HI}}/N_0}}{\sigma_{\rho/\rho_0}} = \left(\frac{\sum_k \ptwod(k)}{\sum_k \pthreed(k)}\right)^{1/2},
    \label{eqn:brunt}
\end{equation}
referred to as the `Brunt Factor', from which we can recover the volume density dispersion itself, $\sigr$.

\subsubsection{Velocity statistics}\label{sssec:vel_stats}
The plane-of-sky dispersion of the velocity centroid can be used to estimate the line-of-sight (LOS) velocity variance, from which we can reconstruct the 3D turbulent velocity dispersion, defined as
\begin{equation}
    \sigv = (\sigma_{v_x}^2 + \sigma_{v_y}^2 + \sigma_{v_z}^2)^{1/2},
\end{equation}
or simply, the sum in quadrature of the velocity variance in each spatial direction $x, y, z$. Of course we cannot measure the variance directly from observations, so instead we assume that the fluctuations in the $x-$ and $y-$directions are the same as the LOS fluctuations. Whilst this assumption may not be reasonable for a cloud trapped in an accelerating flow originating from the GC, we minimise any potential variance by removing systematic motions through our gradient and large-scale subtraction methods. Given the lack of spectral resolution in these data, we cannot directly measure the LOS dispersion (the second moment), and so we make a further assumption that the variance in velocity centroid (first moment) in the plane-of-sky is proportional to the variation along the LOS. To recover $\sigv$ we follow the methods developed by \cite{Stewart:2022ut}, who show that the 3D turbulent velocity dispersion can be recovered from PPV space using the standard deviation of the gradient-corrected moment-1 map together with a correction factor\footnote{This correction factor is the mean of the $p_0$ values in lines~4--6 of Tab.~E1 of \citet{Stewart:2022ut}. We choose the gradient-subtracted statistics and choose the mean of those values, which are independent of the LOS orientation with respect to the rotation axis of the cloud/kernel region.} of $C_\mathrm{(c-grad)}^\mathrm{\,any}=3.3\pm0.5$.

The sonic Mach number ($\mach$) of the turbulent component of the velocity field is given by
\begin{equation}
    \mach = \frac{\sigv}{\cs}, 
    \label{eqn:mach}
\end{equation}
and the sound speed, $\cs$ is defined as
\begin{equation}
    \cs = \left(\frac{\gamma k_\mathrm{B} T}{\mu m_\mathrm{H}}\right)^{1/2},
    \label{eqn:cs}
\end{equation}
where $\gamma = 5/3$ is the adiabatic index, $k_\mathrm{B}$ is the Boltzmann constant, $T$ is the gas temperature, $\mu=1.4$ is the mean particle weight of $\hi$ \citep{KauffmannEtAl2008}, and $m_\mathrm{H}$ is the mass of a hydrogen atom.

\subsubsection{Temperature and sound speed}\label{sssec:temp}
Calculating the Mach number requires us to know the sound speed in the gas, as per Eq.~\ref{eqn:mach}, which in turn requires an estimate of the temperature of the gas. In these particular data, we are constrained by the coarse spectral resolution of 5.5\,km\,s$^{-1}$, which means we cannot directly measure LOS velocity fluctuations smaller than this, and we cannot perform a Gaussian decomposition on the PPV cubes to estimate the fraction of cold gas in these clouds, and therefore estimate an average temperature for the emitting $\hi$. 
In \citetalias{Gerrard:2023aa} we assumed a constant sound speed for the whole SMC, based on the assumption that most of the emitting gas was warm neutral medium (WNM). Here we must consider that the cold gas fraction is likely much more dominant, as there is a relatively large fraction ($\sim 55\%$) of molecular material in C1 (but none in C3) \citepalias{Noon:2023aa}. In this case we use the location of the molecular gas to estimate the temperature variance of the $\hi$ across the clouds, assuming that where the molecular gas dominates the combined (molecular and atomic) column density of the cloud, the spatially corresponding $\hi$ is all cold neutral medium (CNM). In regions where there is $\hi$ without $\hm$, we assume a mixture of the three $\hi$ phases \citep[e.g.][]{Wolfire:2003ug}, with a significant percentage of the gas existing as unstable neutral medium (UNM).

To begin with, we determine how much molecular column density is in each pixel compared to the total column density (atomic plus molecular), i.e., the molecular fraction
\begin{equation}
    f_{\mathrm{mol}} = \frac{N_{\hm}}{N_{\mathrm{HI}} + N_{\hm}}.
\end{equation}

\begin{figure}
     \centering
     \includegraphics[width=\linewidth]{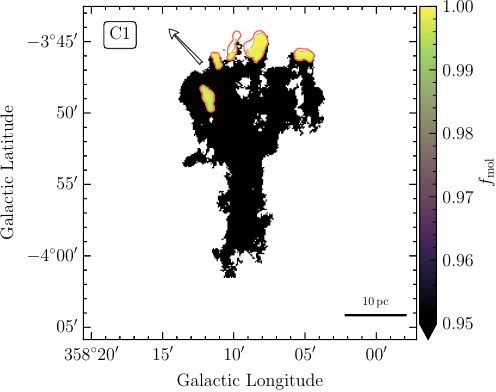}
     \caption{A map of the fraction of molecular column density, $f_{\mathrm{mol}}$ for C1. The colourbar shows the variation of $f_{\mathrm{mol}}$ where there is molecular material, although the amount of $\hm$ only varies between $\sim 0.95$ and 1., meaning that the black regions regions are pure $\hi$ ($f_{\mathrm{mol}} = 0$). The red contours shows $\hm$ column density at $5.0\times10^{21}$\,cm$^{-2}$. The scale bar shows 10\,pc, assuming the physical distance to the clouds is 8.3\,kpc. The arrow in each panel points towards the GC, which is a proxy for the opposite direction in which the Galactic wind is flowing.}
     \label{fig:c1_mol_frac_map}
\end{figure}

Once we have created a map of $f_\mathrm{mol}$ (Fig.~\ref{fig:c1_mol_frac_map}), we need to assign temperatures to each pixel based on the ratio of molecular to atomic gas. Noting that we always observe $\hi$ gas, we need to estimate the $\hi$ temperature, and we do so via the following equation:
\begin{equation}
    T = T_{\mathrm{CNM}}\times f_{\mathrm{mol}} + (1-f_{\mathrm{mol}})\times \overline{T}_{\mathrm{HI}},
    \label{eq:th1}
\end{equation}
where the average $\hi$ temperature ($\overline{T}_{\mathrm{HI}}$) is given by
\begin{equation}
    \overline{T}_{\mathrm{HI}} = \mathrm{f_{CNM}}\times T_{\mathrm{CNM}} + \mathrm{f_{UNM}}\times T_{\mathrm{UNM}} +\mathrm{f_{WNM}}\times T_{\mathrm{WNM}},
    \label{eq:t_avg}
\end{equation}
with the CNM, UNM, and WMN mass fractions and temperatures of each phase listed in Tab.~\ref{tab:hi_phases}. We use the best observational estimates we have for the mass fractions of each phase, taken from the 21-SPONGE survey \citep{Murray:2018aa}. These values are derived from the solar-neighbourhood ISM, and although it is possible that the actual mass fractions of each phase differ from these estimates in the extra-planar clouds, we are working under the assumption that the cloud material was launched from the Galactic plane, resembling the $\hi$ phase composition of the local ISM. Estimates for the temperature range of each phase are taken from \citet{McClure-Griffiths:2023aa}, where theoretical and observational estimates for $T_{\mathrm{CNM}}$, $T_{\mathrm{UNM}}$, $T_{\mathrm{WNM}}$ are collated.

\begin{table}
\centering
\def\arraystretch{1.7}
\caption{Mass fractions and temperatures of CNM, UNM, WNM.}
\begin{tabular*}{\linewidth}{@{\extracolsep{\fill}} llccc }
\hline
\hline
& Phase & CNM & UNM & WNM \\
\hline
(1) & Fraction $f_\mathrm{phase}$ & 0.28 & 0.20 & 0.52 \\
(2) & Minimum $T_{\mathrm{phase}}$ & 25\,K  & 250\,K  & 4000\,K \\
(3) & Maximum $T_{\mathrm{phase}}$ & 250\,K & 4000\,K & 8000\,K \\
\hline
\hline
\end{tabular*}
\label{tab:hi_phases}
\textbf{Notes:} row (1) is the mass fraction for each phase taken from the 21-SPONGE survey \citep{Murray:2018aa}. Rows (2) and (3) are the minimum and maximum temperature for each phase, taken from \citet{McClure-Griffiths:2023aa}.
\end{table}

Eq.~\ref{eq:th1} describes three cases: pixels where there is more $\hm$ than $\hi$ in the column are the CNM temperature, pixels which have a mixture of $\hm$ and $\hi$ have a temperature which is a proportional mixture of $T_{\mathrm{CNM}}$ and the $\overline{T}_{\mathrm{HI}}$, and pixels where there is only $\hi$ have the $\overline{T}_{\mathrm{HI}}$ temperature.

Lastly, we use the minimum and maximum temperatures in Table.~\ref{tab:hi_phases} to construct a minimum and maximum temperature map for C1 via Eq.~\ref{eq:t_avg} and \ref{eq:th1}, and then use Eq.~\ref{eqn:cs} to convert them into maps of the sound speed. We do this because there are several orders of magnitude in the temperature variations, and because the sound speed goes as the square root of the temperature, we are then able to take the arithmetic mean of the minimum and maximum maps as our final map to be used in the analysis, as the sound speed spans a smaller range. Using the lower temperature limits, we find that $\overline{T}_{\mathrm{HI}} \sim 2100$\,K, and with the upper limits $\overline{T}_{\mathrm{HI}} \sim 5000$\,K.

The above prescription is applied to both clouds, but as there is no molecular material in C3 Eq.~\ref{eq:th1} collapses to a constant temperature of $\overline{T}_{\mathrm{HI}}$ (and therefore sound speed) across the cloud.

%%%%%%%%%%%%%%%%%%%%%%%%%%%%%%%%%%%%%%%%%%%%%%%%%%

\section{Results}\label{sec:results}
In this section we present the results of the turbulence analysis described in the previous section. We discuss how the mapped turbulence statistics compare to the physical structure and orientation of the clouds, as well as comparing the median values of the maps with the clouds' global turbulence quantities.

\subsection{Main analysis quantities}
\begin{figure*}
     \centering
     \includegraphics[width=\linewidth]{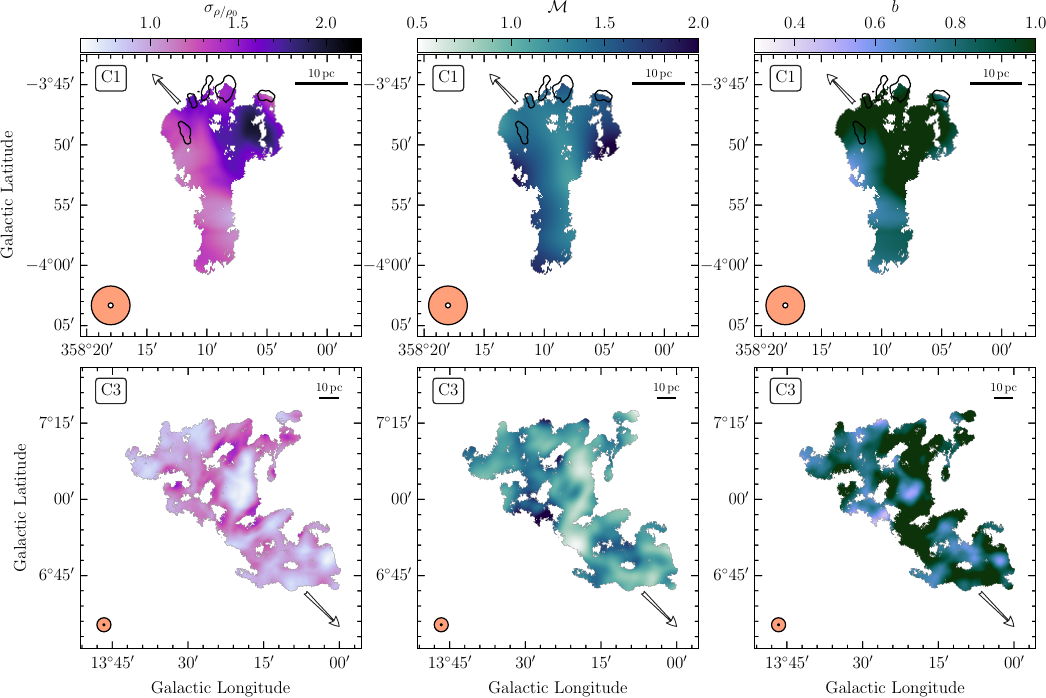}
     \caption{Maps of C1 (top) and C3 (bottom), of the spatially resolved analysis quantities. From left to right, we show the volume density dispersion ($\sigr$), the Mach number ($\mach$) and the turbulence driving parameter ($b$). The orange circle shows the FWHM of the kernel, while the small white circle represents the beam of the \hi{} observations. The black contours show the integrated $\hm$ intensity $5.0\times10^{21}$\,cm$^{-2}$ in C1. The scale bar shows 10\,pc at a distance of 8.3\,kpc, and the arrows point towards the GC. For comparison, the colourbars are constant across clouds.}
     \label{fig:analysis_maps_1}
\end{figure*}
\begin{figure*}
     \centering
     \includegraphics[width=\linewidth]{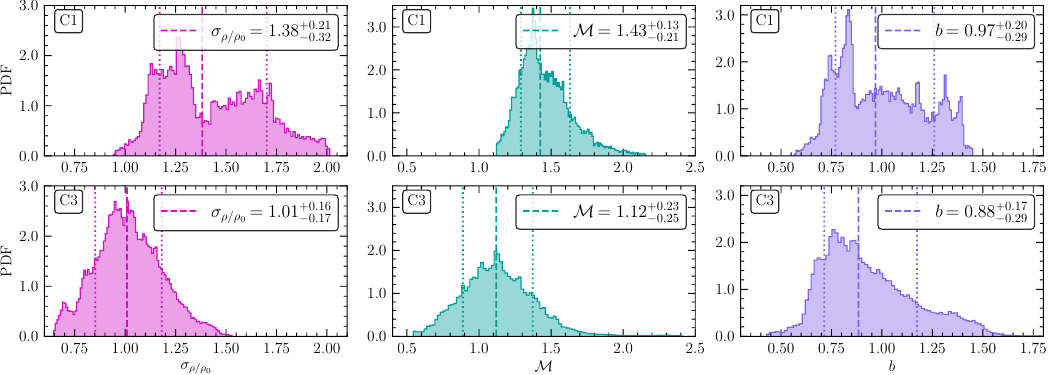}
     \caption{PDFs of the analysis quantities $\sigr$, $\mach$ and $b$ (left to right), for C1 (top) and C3 (bottom). The dashed line shows the median of each distribution, while the dotted lines show the $16^{\mathrm{th}}$ and $84^{\mathrm{th}}$ percentiles, respectively.}
     \label{fig:analysis_pdfs_1}
\end{figure*}
Our methods are designed to measure the volume density dispersion ($\sigr$, Sec.~\ref{sssec:dens_stats}), the Mach number ($\mach$, Sec.~\ref{sssec:vel_stats}) and the turbulence driving parameter ($b$, Eq.~\ref{eq:b}) which we refer to as the main analysis quantities, maps of which are shown in Fig.~\ref{fig:analysis_maps_1}. The first column shows that the variation of $\sigr$ is more pronounced in C3, while C1 has a smoother distribution. C1 has overall higher values of $\sigr$, with a slight correlation between regions of high $\sigr$ and the position of the molecular gas clumps at the `head' of the cloud, when compared to its $\hi$-only `tails'. While we cannot directly map the density itself, higher $\sigr$ means there is a higher chance of regions of large absolute density, possibly signifying a phase transition in the atomic material that is required for molecule formation and maintenance \citep{Krumholz:2009ab, Girichidis:2021aa, Banda-Barragan:2021aa}. 

The Mach number map of C1 shows that there is a trend towards higher $\mach$ on the outside edges of the cloud, while the  Mach number in the inner region is lower. There is some suggestion of higher Mach numbers near the molecular gas, as is to be expected from our construction of the sound speed map, in keeping with colder gas being more supersonic than warm gas. In C3, we have a great deal of variation in Mach numbers, although there is no obvious trend in said variations in relation to the head or tail of the cloud. If we subscribe to the interpretation of C3 as having been entrained in the wind for much longer than C1 \citepalias{Noon:2023aa}, we would expect it to be far more fragmented and disorganised \citep{Armillotta:2016aa, Zhang:2017aa, Banda-Barragan:2019aa, Schneider:2020aa, Gronke:2020aa}. From the aforementioned simulation work, the initial entrainment phase is marked by a distinct head-tail cloud structure, later evolving into an extended structure aligned with the wind, but where turbulent properties are likely to be homogeneous across the cloud, since there is no net shear with the local environment. This is borne out in our maps, as there is no trend in variations of the three main analysis quantities with position or orientation within the cloud. Overall, C3 is more subsonic than C1, although with pockets of higher Mach. This is expected, as C3 has no detectable molecular gas, and as such, we expect the average gas temperature to be warmer in C3 than in C1, and hence have a lower average Mach number.

Finally, taking the ratio of columns~1 and~2 gives us maps of the turbulence driving parameter (column~3). In C1, we can see that the tail of the cloud is driven more solenoidally than the head. This could be due to shearing at the interface of the atomic cloud material and the ionised wind. Overall however, C1 is compressively driven, particularly in the regions containing molecular gas, as is anticipated given the compression of the head of the cloud by the wind. C3 displays an interesting amount of variation across the cloud between solenoidal, mixed and compressive driving, but the driving does not seem to be correlated with the head or tail of the cloud, as with the other two main quantities. It is possible that the compressive borders between the clumpy structures in C3 are cloudlet-cloudlet collisions.

Fig.~\ref{fig:analysis_pdfs_1} shows the probability distribution functions (PDF) of the Mach number, density dispersion and driving parameter, as mapped in Fig.~\ref{fig:analysis_maps_1}. We see that C1 is distributed above $\sigr = 1.0$, whereas the median value in C3 is 1.01. In the second column, we see that all the Mach numbers in C1 are in the trans- to supersonic regime, whereas C3 has a significant portion of the distribution below $\mach=1$. In the third column, we see that the distribution of the turbulence driving parameter is slightly narrower in C1 as compared to C3, but that in the latter distribution the median value is skewed lower by roughly 10\%. This reflects the larger amount of variation seen in the map of $b$ for C3. C1 and C3 both have significant portions of their distributions above $b=1.0$, which is the upper limit describing purely compressive driving. While this presents as `unphysical', as discussed in \citetalias{Gerrard:2023aa}, these spatial measurements of the driving parameter should be taken as an indication of which parts of the cloud are relatively more or less compressively driven. We discuss the uncertainty associated with our measurement of $b$ quantitatively in Sec.~\ref{ssec:uncertainties}.

\subsection{Intermediate analysis quantities}
\begin{figure*}
     \centering
     \includegraphics[width=\linewidth]{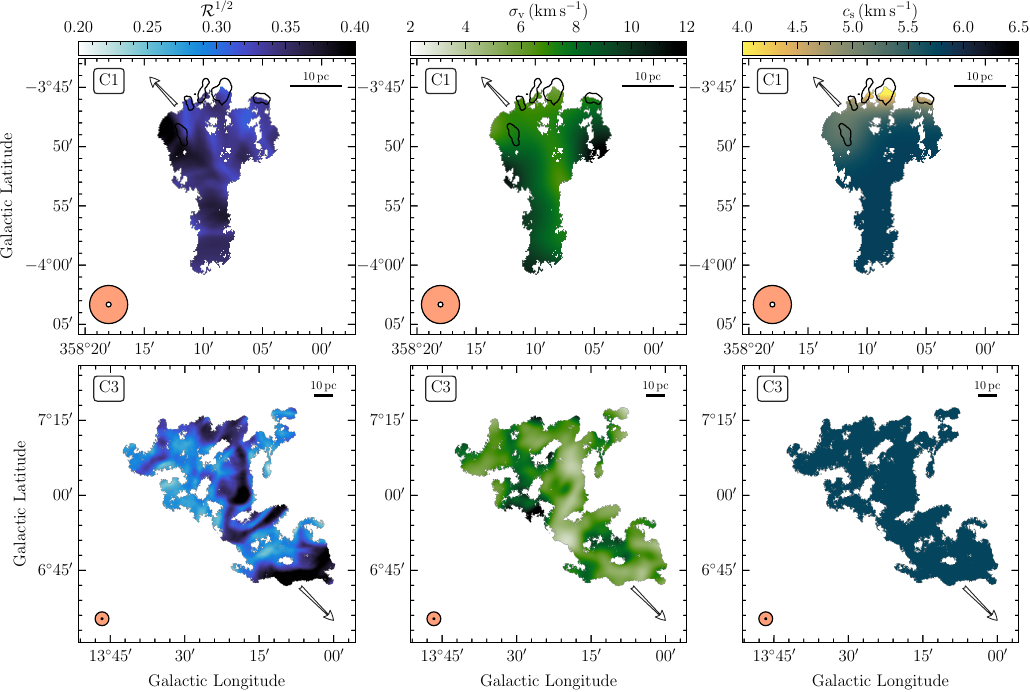}
     \caption{Same as Fig.~\ref{fig:analysis_maps_1} but showing the Brunt Factor ($\brunt$), the 3D velocity dispersion ($\sigv$) and the sound speed $\cs$. Because there is no molecular gas in C3, the sound speed is a constant value everywhere in this map.}
     \label{fig:analysis_maps_2}
\end{figure*}
\begin{figure*}
     \centering
     \includegraphics[width=\linewidth]{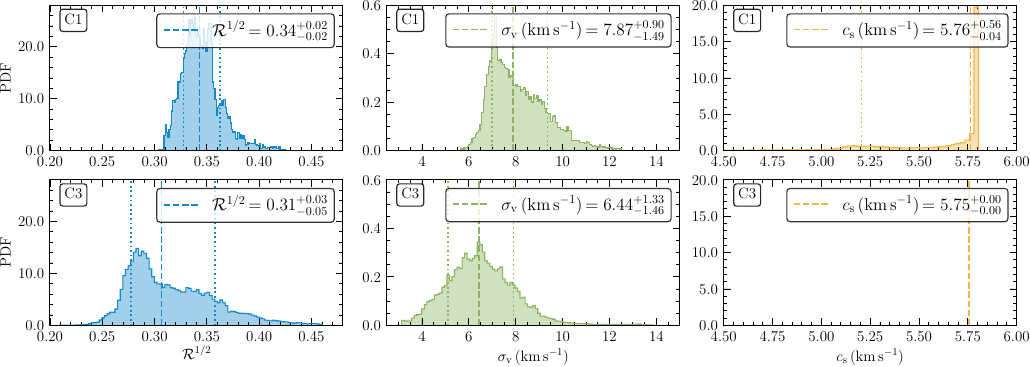}
     \caption{Same as Fig.~\ref{fig:analysis_pdfs_1} but showing the intermediate analysis $\brunt$, $\sigv$ and $\cs$.}
     \label{fig:analysis_pdfs_2}
\end{figure*}

Fig.~\ref{fig:analysis_maps_2} shows the maps of the intermediate analysis quantities that are needed to construct the main analysis quantities, but that are nonetheless important in their own right. In the first column we see that the Brunt Factor ($\brunt$) is predominantly of order $0.3 \sim 0.4$ for both cloud, which is what we expect from previous findings \citep{Brunt2010,Ginsburg:2013ww,Federrath:2016ac,Menon:2021vj,Sharda:2022vg,Gerrard:2023aa}. Comparing the map of the Brunt factor to $\sigr$ in Fig.~\ref{fig:analysis_maps_1}, we see that the variations in $\brunt$ do not dominate the variations in $\sigr$, such that the volume density dispersion is primarily a consequence of the observed column density. We discuss the errors introduced by the Brunt Method further in Sec.~\ref{ssec:uncertainties}. 

The variation in the 3D velocity field is quite smooth across C1, although it exhibits higher $\sigv$ towards the edges of the tail of the cloud, which is reflected in the Mach number map in Fig.~\ref{fig:analysis_maps_1}. It appears that there is a slight trend towards lower $\sigv$ in the molecular regions. C3 has the lowest $\sigv$ values of the two clouds, with a similar amount of variation across the cloud as in its $\sigr$ map, which is again reflected in the $\mach$ map.
By construction, the sound speed map (column~3) of C1 has the lowest values closest to where the molecular clumps are. The map has been processed through the same kernel method as all the other quantities, which has smoothed out any steep transitions between high and low sound speeds. Of course the sound speed of C3 is a constant value as it has no molecular component, and as such our simple model for the temperature does not include any variation in this case.

Fig.~\ref{fig:analysis_pdfs_2} shows PDFs of the intermediate analysis quantities. C3 has a broader distribution of Brunt factors than C1, and an overall lower median value. C1 has a larger value of $\sigv$ than in C3, as well as a more asymmetrical distribution. Dividing $\sigv$ by the sound speed (third column) gives us the Mach number in Fig.~\ref{fig:analysis_pdfs_1}. By construction, C3 has only one value for the sound speed, while C1 has a range of values but the bulk of the cloud is at the $\hi$ average sound speed (as outlined in Sec.~\ref{sssec:temp}).

\subsection{Uncertainties}\label{ssec:uncertainties}
There are several sources of systematic error associated with our methods. The error associated with the Brunt method is quoted as $\sim 10\%$ in \citet{BruntFederrathPrice2010a}, when the Fourier image of the column density is axisymmetric. In Appendix~\ref{app:brunt}, we discuss the level of anisotropy in each cloud and conclude that the upper limit on the error introduced via the Brunt method is 40\% in this instance.

Converting the centroid velocity variance to $\sigv$ via the Stuart method introduces an error of $\sim 10\%$ \citep{Stewart:2022ut}, which is carried through to our estimates of the Mach number. Lastly, we must consider the error introduced by our temperature method. For this, we will take the upper and lower bound on the possible values that the temperature and therefore $\cs$ can take depending on the temperature ranges of each $\hi$ phase, as described in Sec.~\ref{sssec:temp}. For C1, the mean of the sound speed map constructed using the maximum temperatures of each phase is 6.7\,km\,s$^{-1}$, and the mean of the same map constructed with the minimum temperatures of each phase is 4.4\,km\,s$^{-1}$. For C3 we have 7.0\,km\,s$^{-1}$ and 4.6\,km\,s$^{-1}$ respectively. Comparing these values to the sound speeds derived using the mean of the minimum and maximum temperature maps gives us a relative error on the global sound speed of both maps of $\sim 20\%$. 

We collect the relative errors outlined above and propagate them through the global quantities for each cloud. This results in a relative error of $\sim 22\%$ for the Mach number, and $\sim 46\%$ for the driving parameter. Table~\ref{tab:physical_quantities} summarises the bulk characteristics of each cloud, like $\hi$ mass, molecular mass, age and distance from the galactic centre, all from \citetalias{Noon:2023aa}. To add to this, we list the median values of the mapped quantities, with error bars representing the $\minerr$ and $\maxerr$, as well as the global values for each cloud which we compute directly on the LPF-corrected moment-0 and moment-1 maps. We find that when comparing the mapped and global quantities, the latter is lower than the former across all values in C1, but the opposite is true (with the exception of $\brunt$) for C3. This is likely due to the size of the analysis kernel compared to the cloud in each case, as C3 is a much larger object and therefore the smallest resolvable kernel is small compared to its on-sky extent. This means that the scale probed by the mapped quantities is smaller by 1-2 orders of magnitude (depending on the cloud) than that probed by the global quantities. However, comparing the clouds and the scales on which the analysis quantities are calculated on, all the values agree within $1\sigma$, except for $\sigv$ in C3. Theoretically, the velocity variance should scale with the length scale on which it is measured as a power law with an exponent of $p\sim0.5$ in the supersonic regime, and $p\sim 0.4$ in the subsonic regime \citep[e.g.][]{Larson:1981aa, Ossenkopf:2002aa, Heyer:2004aa, Federrath:2021wr}, so it is expected that the global quantities of the clouds are higher than median values measured on the kernel scale.

\begin{table}
\centering
\def\arraystretch{1.7}
\caption{Summary of key quantities for C1 and C3. The top section of the table shows the \emph{kinematic wind model quantities} (age and distance from the GC) from \citeauthor{Di-Teodoro:2018aa} and \citeauthor{Lockman:2020aa}, assuming a distance to the GC of 8.2 kpc and the subsequent \emph{derived masses} ($\hi$ and $\hm$ masses) from \citetalias{Noon:2023aa}. The \emph{derived masses} are calculated using an arbitrary boundary of each cloud, both spectrally and spatially as described in \citetalias{Noon:2023aa}. The middle section shows the median analysis quantities, with error bars of the $\minerr$ and $\maxerr$ percentiles of each distribution. The bottom section shows the analysis quantities calculated for the entire cloud, with error bars as outlined in Sec.~\ref{ssec:uncertainties}.}
\begin{tabular*}{\linewidth}{@{\extracolsep{\fill}} lccl }
\hline
\hline
& C1 &  C3 & Reference\\
\hline
\emph{Kinematic model} &&&\citeauthor{Di-Teodoro:2018aa},\\ \emph{quantities}&&&\citeauthor{Lockman:2020aa}\\
\hline
Distance (kpc) & 0.8 & 2.1 & \\
Age (Myr) & 3.5  & 8.3 & \\
\hline
\emph{Derived masses} &&& \citetalias{Noon:2023aa}\\
\hline
M$_{\hi}\,(\msun)$ & 385  & 3426 & \\
M$_{\hm}\,(\msun)$ & 443  & - & \\
\hline
\emph{Mapped quantities}&&& This work \\
\hline
$\mathcal{R}^{1/2}$                    & $0.34^{+0.02}_{-0.02}$  & $0.31^{+0.03}_{-0.05}$ \\
$\sigr$                                & $1.4^{+0.2}_{-0.3}$  & $1.0^{+0.2}_{-0.2}$ \\ 
$\sigma_{v,\mathrm{3D}}$\,(km\,s$^{-1}$)       & $7.9^{+0.9}_{-1.5}$  & $6.4^{+1.3}_{-1.5}$ \\
$\cs$\,(km\,s$^{-1}$)                          & $5.8^{+0.6}_{-0.04}$  & $5.8^{+0.0}_{-0.0}$ \\
$\mach$                                & $1.4^{+0.1}_{-0.2}$  & $1.1^{+0.2}_{-0.3}$ \\
$b$                                    & $1.0^{+0.2}_{-0.3}$  & $0.9^{+0.2}_{-0.3}$ \\
\hline
\emph{Global quantities}&&& This work\\
\hline
$\mathcal{R}^{1/2}$              & $0.4 \pm 0.1$  & $0.2 \pm 0.1$ \\
$\sigr$                          & $1.5 \pm 0.6$  & $1.8 \pm 0.7$ \\ 
$\sigma_{v,\mathrm{3D}}$\,(km\,s$^{-1}$ & $9.6 \pm 1.0$  & $11.0 \pm 1.1$ \\
$\cs$\,(km\,s$^{-1}$                    & $5.5 \pm 1.1$  & $5.8 \pm 1.2$ \\
$\mach$                          & $1.7 \pm 0.4$  & $2.0 \pm 0.4$ \\
$b$                              & $0.9 \pm 0.4$  & $0.9 \pm 0.4$ \\
\hline
\hline
\end{tabular*}
\label{tab:physical_quantities}
\end{table}

\subsection{Context}
\begin{figure*}
     \centering
     \includegraphics[width=\linewidth]{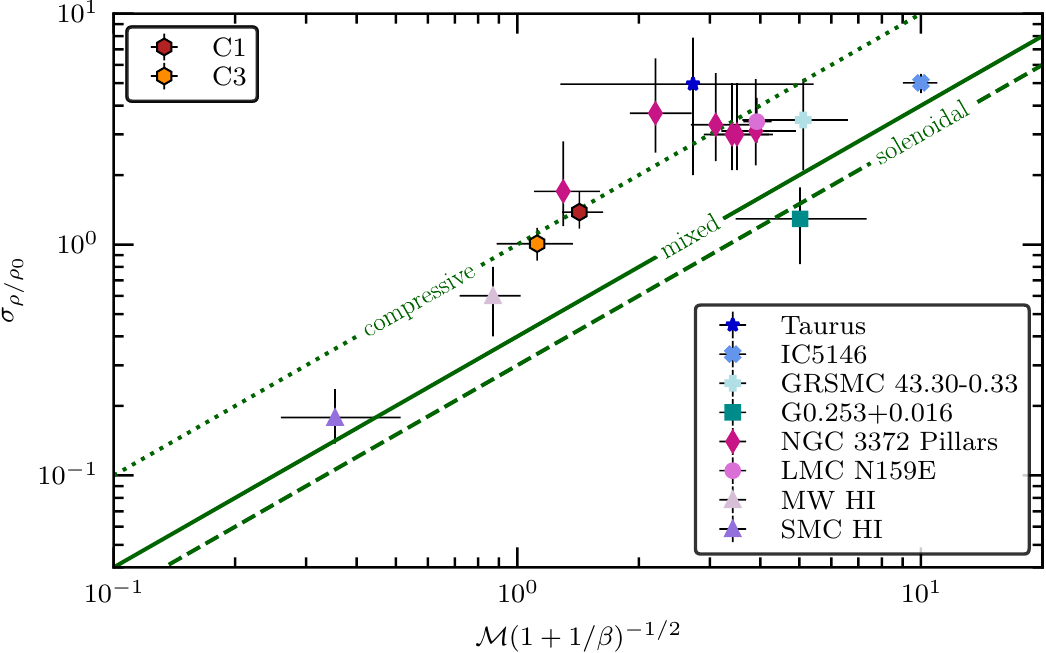}
     \caption{A summary of the available observational estimates for the density dispersion -- Mach number relation in different environments. The y-axis shows the volume density dispersion ($\sigr$), and the x-axis shows the turbulent Mach number ($\mach$), including a factor involving plasma $\beta$ (ratio of thermal to magnetic pressure), as some of the literature values shown (Taurus and G0.253+0.016) have been calculated using the magnetised version of the $\sigr-\mach$ relation in Eq.~\ref{eq:b} \citep[see][]{Molina:2012uv}. The three diagonal lines show the theoretical limits for compressive ($b=1.0$, dotted), mixed ($b=0.38$, solid) and solenoidal ($b=0.33$, dashed) driving of the turbulence \citep{Federrath:2010wz}. The hexagons show the median values for C1 (red) and C3 (orange). The error bars on these points show the $16^{\mathrm{th}}$ to $84^{\mathrm{th}}$ percentile on each axis. For context we include a variety of sources from the literature: Taurus (dark blue star) \citep{Brunt2010}, which includes magnetic field estimates and revised Mach number estimations from \citet{Kainulainen:2013wh}, using $^{13}$CO line imaging observations; IC5146 (blue cross) \citep{PadoanJonesNordlund1997}, using $^{12}$CO and $^{13}$CO observations; GRSMC~43.30-0.33 (aqua plus) \citep{Ginsburg:2013ww}, observed in H$_{2}$CO absorption and $^{13}$CO emission; `The~Brick' (G0.253+0.016, teal square) \citep{Federrath:2016ac}, using HNCO observations; `The Pillars of Creation' (NGC~3372 pillars, magenta diamonds) \citep{Menon:2021vj}, from $^{12}$CO, $^{13}$CO and C$^{18}$O; `The Papillon Nebula' (LMC N159E, pink circle) \citep{Sharda:2022vg}, again in $^{12}$CO, $^{13}$CO and C$^{18}$O; the WNM in the MW (lilac triangle) \citep{Marchal:2021tz} ($\hi$ observations); and the WNM in the SMC (purple triangle) (\citetalias{Gerrard:2023aa}).}
     \label{fig:context}
\end{figure*}

As outlined in the introduction, there have been several measurements of the turbulence driving parameter from observational data, mostly in star-forming molecular clouds. Fig.~\ref{fig:context} summarises these previous measurements, with the addition of C1 and C3, showing that these clouds sit between previous measurements of the diffuse $\hi$ and the dense molecular clouds on the $\sigr-\mach$ diagram. As we have shown, C3 has a lower Mach number and density dispersion, and so sits below C1 in the sub-to-transonic regime on this figure. More high-resolution observations of these extra-planar $\hi$ clouds will allow us to further populate this figure with these kind of objects, but from this investigation it seems that $\hi$ clouds of this type have lower $\mach$ and $\sigr$ values than their molecular counterparts, although the ratio still results in high levels of compressive driving which is comparable to the molecular pillars in NGC~3372 \citep{Menon:2021vj}.
%%%%%%%%%%%%%%%%%%%%%%%%%%%%%%%%%%%%%%%%%%%%%%%%%%

\section{Summary} \label{sec:summary}
In this study, we employ the \citetalias{Gerrard:2023aa} methods for calculating the properties of the internal turbulence of two extra-planar \hi{} clouds, and develop a novel approach for modelling the spatially varying temperature of a mixture of atomic and molecular hydrogen. Cloud~C1  contains a significant amount ($\sim55$\,\% by mass) of molecular material, is small and compact, at $b\sim-3.9^{\circ}$ below the disc. Cloud~C3 is much larger and more spatially dispersed, it contains no detectable molecular material, and is at a latitude of $b\sim7.0^{\circ}$ above the disc. Despite their physical differences and distance from the plane of the Milky Way, the internal turbulence statistics of these clouds are very similar. We find that both clouds are in the sub-to-trans-sonic regime with $\mach ~\sim 1.4^{+0.1}_{-0.2}$ in C1 and $\mach ~\sim 1.1^{+0.2}_{-0.3}$ in C3. We find that the turbulence in both clouds is predominately compressively-driven with $b ~\sim 1.0^{+0.2}_{-0.3}$ in C1 and $b ~\sim 0.9^{+0.2}_{-0.3}$ in C3. The volume density contrast in C1 is $\sigr \sim 1.4^{+0.2}_{-0.3}$ and in C3 $\sigr \sim 1.0^{+0.2}_{-0.2}$. Comparing the median of the spatially varying analysis quantities and the global quantities for each cloud, there is good $1\sigma$ agreement across all quantities and clouds. We find that there are no obvious trends in the spatial variation of the analysis quantities with orientation or position of C3, but C1 shows signs of the cloud-wind interaction at the head of the cloud leading to more compressively-driven turbulence in that region. This is consistent with the idea that the wind compressing the tip of the clouds drives compressive turbulence, but that this effect of the cloud-wind interaction may become less focused as the cloud disperses during its journey away from the Galactic disc.

%%%%%%%%%%%%%%%%%%%%%%%%%%%%%%%%%%%%%%%%%%%%%%%%%%
\section*{Acknowledgements}
The authors acknowledge Interstellar Institute’s program “II6” and the Paris-Saclay University’s Institut Pascal for hosting discussions that nourished the development of the ideas behind this work. The MeerKAT telescope is operated by the South African Radio Astronomy Observatory, which is a facility of the National Research Foundation, an agency of the Department of Science and Innovation. I.A.G.~would like to thank the Australian Government and the financial support provided by the Australian Postgraduate Award. 
C.F.~acknowledges funding by the Australian Research Council (Discovery Projects grant~DP230102280), and the Australia-Germany Joint Research Cooperation Scheme (UA-DAAD). C.F.~further acknowledges high-performance computing resources provided by the Leibniz Rechenzentrum and the Gauss Centre for Supercomputing (grants~pr32lo, pr48pi and GCS Large-scale project~10391), the Australian National Computational Infrastructure (grant~ek9) and the Pawsey Supercomputing Centre (project~pawsey0810) in the framework of the National Computational Merit Allocation Scheme and the ANU Merit Allocation Scheme, through which the data analyses presented in this paper were performed. E.D.T.~was supported by the European Research Council (ERC) under grant agreement no.~10104075. This research was partially funded by the Australian Government through an Australian Research Council Australian Laureate Fellowship (project number~FL210100039) to N.Mc-G.

%%%%%%%%%%%%%%%%%%%%%%%%%%%%%%%%%%%%%%%%%%%%%%%%%%
\section*{Data Availability}
The data underlying this article along with a general implementation of the code used to process the data cubes is available via Zenodo at https://doi.org/10.5281/zenodo.8060960.
\bibliographystyle{mnras}
\bibliography{gerrard, federrath, noon} 

%%%%%%%%%%%%%%%%% APPENDICES %%%%%%%%%%%%%%%%%%%%%
\appendix

\section{Kernel Size} \label{app:kernel_size}
The size of the kernel FWHM is set as the diameter of a circle with an area 3 times the minimum viable area required to recover the statistics in each kernel window. Limiting the size of the kernel as much as possible so that we recover as much spatial detail as possible must be balanced with making the kernel large enough so that the edges are not eaten away during the processing, and we recover as much of the area covered by the raw input data as possible. We explored this for C1, as shown in Fig.~\ref{fig:area_factor} below, comparing the area covered by the input and output maps, depending on how large we made the kernel in comparison to the minimum viable area. We see that by making the kernel 3 times as large as the minimum area we recover 95\% of the input data, and that larger areas don't drastically increase the amount of area recovered. For each cloud, therefore, we make the FWHM of the kernel such that the area contained within it is 3 times the minimum viable area, which is to say that for any given window, at least a third of the area inside the kernel FWHM must be viable data. This is all ultimately set by the beam size, as we are ensuring that at least 20 beams worth of area is contained in this region. The major axis of the beam for each PPV cube is about $\sim25"$.
\begin{figure}
     \centering
     \includegraphics[width=\linewidth]{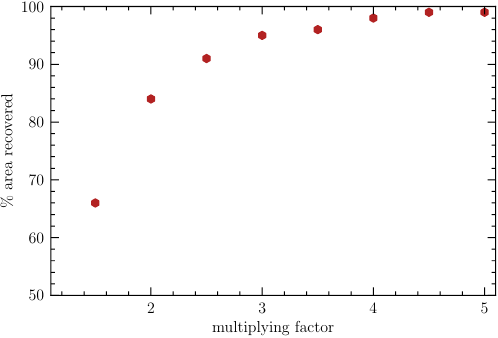}
     \caption{The percentage area of the raw data recovered by the pipeline for C1. Using a multiplying factor of 3 recovers 95\% of the raw data, which is what we have chosen to use throughout the analysis to set the kernel FWHM.}
     \label{fig:area_factor}
\end{figure}

\section{Low-Pass Filtering method} \label{app:LPF}
\begin{figure*}
     \centering
     \includegraphics[width=\linewidth]{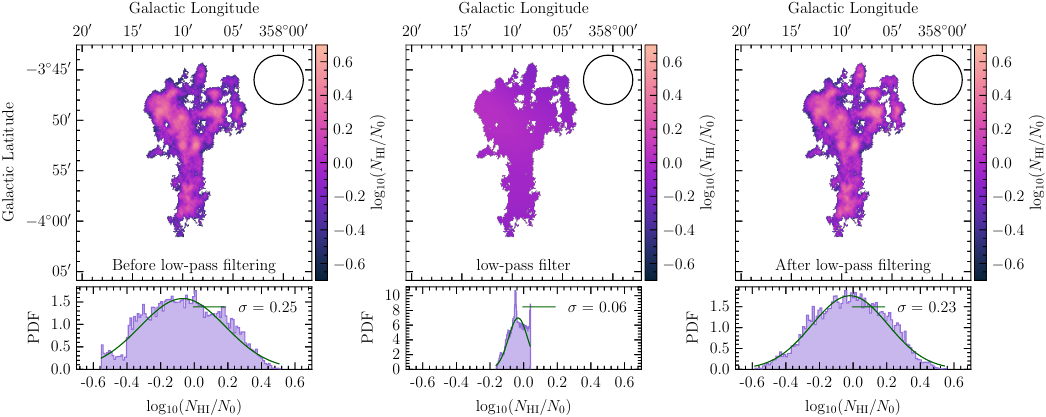}
     \caption{C1 column density. The left-hand panel shows the log of the column density map prior to background subtraction above, and a PDF of the column density below. The middle panel shows the map of the low-pass background filter and its PDF, while the right-hand panel shows the background-subtracted map. On each PDF, a fitted Gaussian is shown in green, along with its $\sigma$ value. The circle in the top right-hand corner of each panel represents the FWHM of the convolution kernel.}
     \label{fig:c1_column_density}
\end{figure*}

\begin{figure*}
     \centering
     \includegraphics[width=\linewidth]{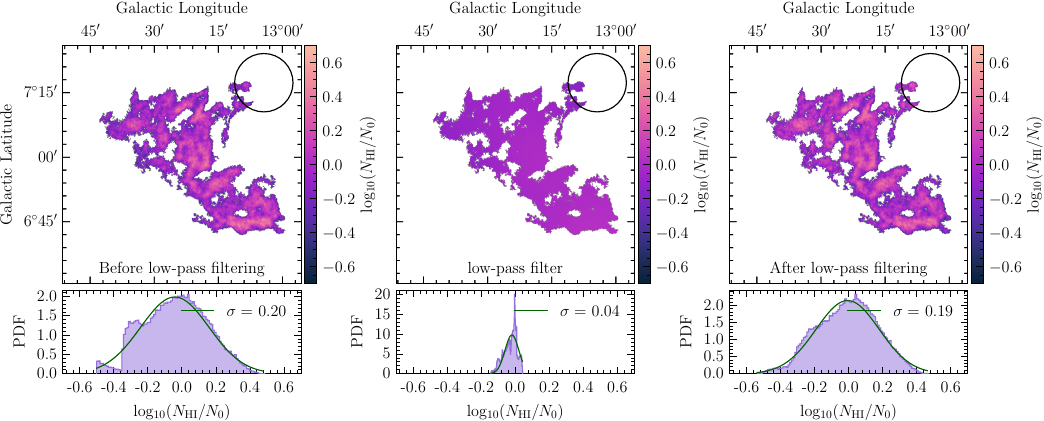}
     \caption{Same as Fig.~\ref{fig:c1_column_density}, but for C3.}
     \label{fig:c3_column_density}
\end{figure*}

As discussed in Sec.~\ref{sssec:LPF}. to extract 3D density statistics from the column density maps, we first subtract the low-pass filter from each map, such that the resulting column density map approximates a log-normal distribution. The results of this are shown in Fig.~\ref{fig:c1_column_density} and~\ref{fig:c3_column_density}. 

\begin{figure*}
     \centering
     \includegraphics[width=\linewidth]{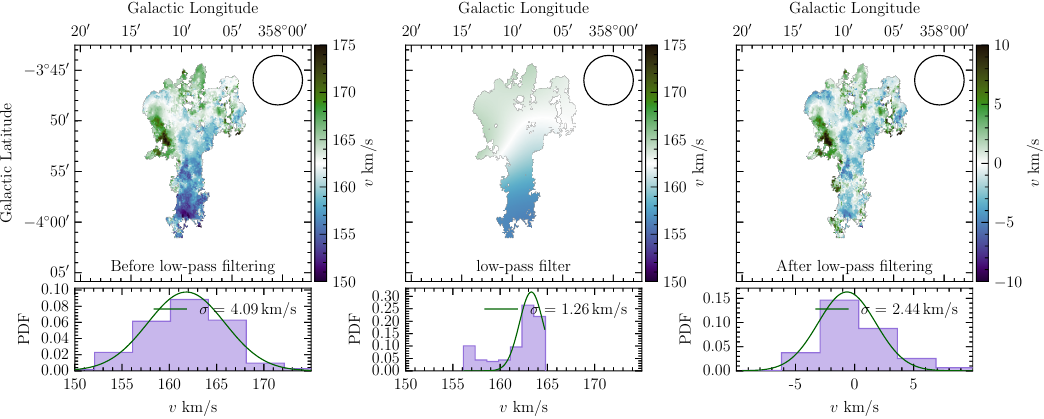}
     \caption{C1 centroid velocity. The panels represent the same steps as in Fig.~\ref{fig:c1_column_density}, and the overlays are the same.}
     \label{fig:c1_centroid_velocity}
\end{figure*}

We apply the LPF-method to the centroid velocity maps in the same way as the column density, which can be seen in Fig.~\ref{fig:c1_centroid_velocity} and~\ref{fig:c3_column_density}. 
\begin{figure*}
     \centering
     \includegraphics[width=\linewidth]{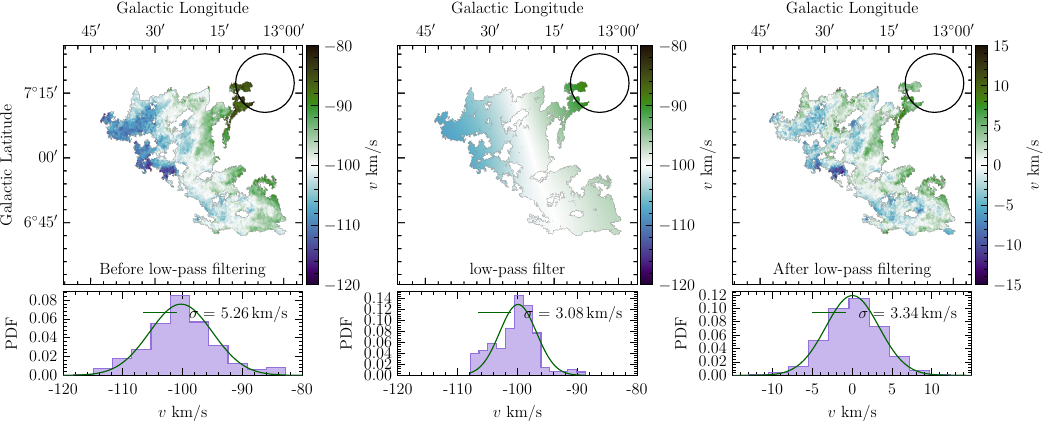}
     \caption{Same as Fig.~\ref{fig:c1_centroid_velocity}, but for C3.}
     \label{fig:c3_centroid_velocity}
\end{figure*}

\section{Brunt method \& anisotropy in column density power spectra} \label{app:brunt}
\begin{figure*}
     \centering
     \includegraphics[width=\linewidth]{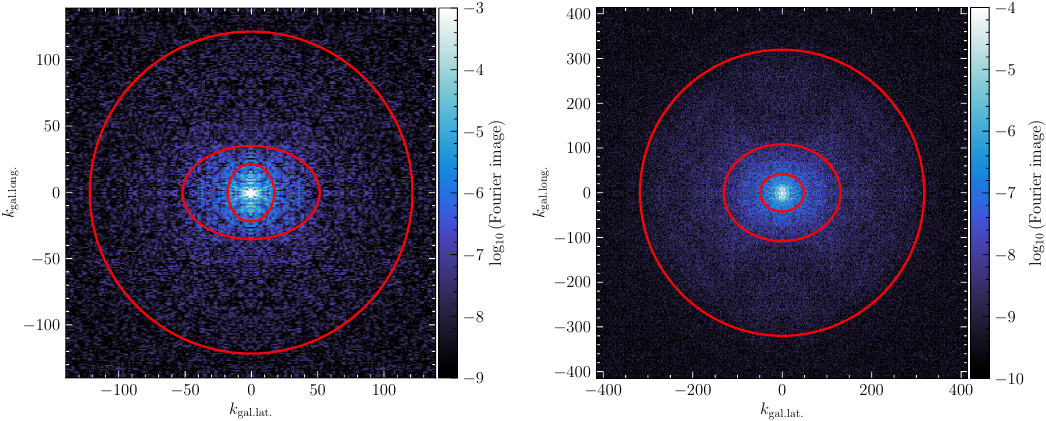}
     \caption{2D power spectra of the normalised column density of the two clouds. \emph{Left:} C1 with contours fitted at $10^{-5}$, $10^{-6}$, $10^{-7}$ corresponding to ellipticities of $e=1.04,\,1.17,\,1.20$. \emph{Right:} C3 with contours fitted at $10^{-4}$, $10^{-5}$, $10^{-6}$ corresponding ellipticities of $e=1.24,\,1.47,\,1.00$.}
     \label{fig:pspec}
\end{figure*}

The Brunt method, described in Sec.~\ref{sssec:dens_stats}, relied on the assumption of the column density fluctuations being isotropically distributed in $k$-space. To ascertain the amount of uncertainty introduced by using this method to obtain $\sigr$, we fit ellipses to the Fourier image of C1 and C3 (Fig.~\ref{fig:pspec}) to investigate the symmetry in the 2D power spectra. In C1, the largest ellipticity is $e=1.47$, while in C3 it is $e=1.21$, which translates to a 47\% deviation from circular in C1 and a 21\% deviation in C3. This deviation can be used as a proxy for the amount of anisotropy in the images. Following \citet{Federrath:2016ac}, we conclude that the error associated with the conversion of the 2D-to-3D density contrast conversion is of order $<40\%$.

%%%%%%%%%%%%%%%%%%%%%%%%%%%%%%%%%%%%%%%%%%%%%%%%%%
% Don't change these lines
\bsp	% typesetting comment
\label{lastpage}
\end{document}